\documentclass[12pt]{article}
\usepackage{amsfonts}
\newcommand{\sfrac}[2]{{\textstyle{\frac{#1}{#2}}}}
\newcommand{\ZZ}{{\mathbb Z}}

\begin{document}
\newpage
\pagestyle{empty}
\setcounter{page}{0}
\vfill
\begin{center}

{\Large {\bf A crystal base for the genetic code}}

\vspace{10mm}

{\large L. Frappat, P. Sorba}

\vspace{4mm}

{\em Laboratoire de Physique Th\'eorique ENSLAPP, URA 1436,}
\\
{\em \'Ecole Normale Sup\'erieure de Lyon and Universit\'e de Savoie, 
France}

\vspace{7mm}

{\large A. Sciarrino}

\vspace{4mm}

{\em Dipartimento di Scienze Fisiche, Universit\`a di Napoli
       ``Federico II''}
\\
{\em and I.N.F.N., Sezione di Napoli, Italy}
\end{center}

\vfill

\begin{abstract}
The quantum enveloping algebra ${\cal U}_{q}(sl(2) \oplus sl(2))$ in the 
limit $q \rightarrow 0$ is proposed as a symmetry algebra for the 
genetic code.  In this approach the triplets of nucleotids or codons in 
the DNA chain are classified in crystal bases, tensor product of ${\cal 
U}_{q \rightarrow 0}(sl(2) \oplus sl(2))$ representations.  Such a 
construction might be compared to the baryon classification from quark 
building blocks in elementary particles physics, one of the main 
differences standing in the property of a crystal base to provide a 
natural order in the state constituents, this order being crucial in the 
codon. Then an operator ensuring the correspondence codon/amino-acid 
can be constructed out of the above algebra. It will be called the 
reading operator, and be such that two codons relative to the same (resp. 
different) amino-acid(s) acquire the same (resp. different) eigenvalue(s).
\end{abstract}

\def\abstractname{R\'esum\'e}
\begin{abstract}
L'alg\`ebre enveloppante quantique ${\cal U}_{q}(sl(2) \oplus sl(2))$ 
dans la limite $q \rightarrow 0$ est propos\'ee comme alg\`ebre de 
sym\'etrie du code g\'en\'etique. Dans cette approche, les triplets de 
nucl\'eotides ou codons dans la chaine d'ADN sont classifi\'es dans 
des bases cristallines, produit tensoriel de repr\'esentations 
de ${\cal U}_{q \rightarrow 0}(sl(2) \oplus sl(2))$. Une telle 
construction peut \^etre compar\'ee \`a la classification des baryons 
\`a partir des quarks en physique des particules \'el\'ementaires, une 
des diff\'erences essentielles r\'esidant dans la propri\'et\'e d'une base 
cristalline de fournir un ordre naturel des constituents, cet ordre 
\'etant crucial dans le codon. Nous construisons un op\'erateur assurant 
la correspondance codon/acide amin\'e, appel\'e op\'erateur de lecture. 
Cet op\'erateur est tel que deux codons relatifs au m\^eme (resp. 
\`a des diff\'erents) acide(s) amin\'e(s) ont des valeurs propres 
identiques (resp. diff\'erentes).
\end{abstract}

\vfill
\vfill

\rightline{ENSLAPP-AL-671/97}
\rightline{DSF-97/37}
\rightline{physics/9801027}
\rightline{December 1997}

\newpage
\pagestyle{plain}
\baselineskip=18pt


The mystery of the perfect correspondence between triplets of 
nucleotides or codons in the desoxyribonucleic acid (DNA) sequence and 
the amino-acids is called the genetic code \cite{livre}.  Let us, in a 
few words, remind how the DNA conducts the synthesis of proteins, which 
constitute the most abundant organic substances in living matter 
systems.  Indeed, the DNA macromolecule is made of two linear chains 
of nucleotides in the famous double helix structure.  Each nucleotide 
is characterized by one of the four elementary bases: adenine (A) and 
guanine (G) deriving from purine, and cytosine (C) and thymine (T) 
coming from pyrimidine.  The DNA is localized in the nucleus of the 
cell and the transmission of the genetic information in the cytoplasm 
is achieved by the messenger ribonucleic acid or mRNA. This operation 
is called the transcription, the A, G, C, T bases in the DNA being 
respectively associated to the U, C, G, A bases, U denoting the 
uracile base.  Finally, a codon is an ordered sequence of three bases 
(e.g.  AAG, ACG, etc.)  and it is a simple exercise to numerate $4 
\times 4 \times 4$ different codons.  Except the three following 
triplets UAA, UAG and UGA, each of the 61 others is related through a 
ribosome to an amino-acid in the universal eukariotic code (see Table 
\ref{table2}).  Thus the chain of nucleotides 
in the mRNA -- and also in the DNA -- can also be viewed as a sequence 
of triplets, each corresponding to an amino-acid, except the three 
above mentioned ones.  These last codons are called non-sense or 
stop-codons, and their role is to stop the biosynthesis.

One can distinguish 20 different amino-acids 
\footnote{They are denoted by the letters Ala (Alanine), Arg (Arginine), 
Asn (Asparagine), Asp (Aspartic acid), Cys (Cysteine), Gln (Glutamine), 
Glu (Glutamic acid), Gly (Glycine), His (Histidine), Ile (Isoleucine), 
Leu (Leucine), Lys (Lysine), Met (Methionine), Phe (Phenylalanine), 
Pro (Proline), Ser (Serine), Thr (Threonine), Trp (Tryptophane), 
Tyr (Tyrosine), Val (Valine).}.  
It follows that different codons can be associated to the same
amino-acid, or in other words, that the genetic code is degenerated. 
Considering the standard eukariotic code (see Table \ref{table2}),
one remarks that the codons are organized in sextets, quadruplets,
triplets, doublets and even singlets, each multiplet corresponding to
a specific amino-acid.  Such a picture naturally led Hornos and
Hornos \cite{Hornos} to look for an underlying symmetry based on a
continuous Lie group.  More precisely, the authors tried to answer
the following question: is it possible to determine a Lie group $G$
carrying a 64-dimensional irreducible representation $R$ and
admitting a subgroup $H$ such that the decomposition of $R$ into
irreducible representations under $H$ gives exactly the different
just above mentioned multiplets?  They proposed as starting symmetry
the symplectic group $Sp(6)$ with successive breakings up to its
Cartan part $U(1) \times U(1) \times U(1)$.

Interpreting the double origin of the nucleotides, each arising either 
from purine or from pyrimidine, as a $\ZZ_{2}$-grading, a 
supersymmetric extension of the above model has been proposed 
\cite{BTJ} with the superalgebra $sl(6 \vert 1)$ as the classification 
(super)algebra before symmetry breaking. A systematic search for 
superalgebras the representation theory of which comes close to the 
multiplet structure of the genetic code has also been recently carried 
out in ref. \cite{SF}.

It is a rather different point of view that we will adopt in this 
letter.  Indeed we will consider the four nucleotides as basic states of 
the $(\sfrac{1}{2},\sfrac{1}{2})$ representation of the ${\cal 
U}_{q}(sl(2) \oplus sl(2))$ quantum enveloping algebra in the limit $q 
\rightarrow 0$.  Then a triplet of nucleotides will be obtained by 
constructing the tensor product of three such four dimensional 
representations.  Actually, this approach mimicks the group theoretical 
classification of baryons made out from three quarks in elementary 
particles physics, the building blocks being here the A, C, G, T (U) 
nucleotides.  The main and essential difference stands in the property of 
a codon to be an {\em ordered} set of three nucleotides, which is not the 
case for a baryon.  Let us be more explicit on an example: there are 
three different codons made of the A, A, U nucleotides, namely AAU, AUA 
and UAA, while the proton appears as a weighted combination of the two 
$u$ quarks and one $d$ quark, that is $\vert p \rangle \sim \vert uud 
\rangle + \vert udu \rangle + \vert duu \rangle$, where the spin 
structure is implicit.

Constructing such pure states is made possible in the framework of any 
algebra ${\cal U}_{q \rightarrow 0}({\cal G})$ with ${\cal G}$ being any 
(semi)-simple classical Lie algebra owing to the existence of a special 
basis, called crystal basis, in any (finite dimensional) representation 
of ${\cal G}$.  The algebra ${\cal G} = su(2) \oplus su(2) \simeq so(4)$ 
appears the most natural for our purpose.  First of all, it is 
``reasonable'' to represent the four nucleotides in the fundamental 
representation of ${\cal G}$.  Moreover, the complementary rule in the 
DNA--mRNA transcription may suggest to assign a {\em quantum number} 
with opposite values to the couples (A,T/U) and (C,G).  The distinction 
between the purine bases (A,G) and the pyrimidine ones (C,T/U) can be 
algebraically represented in an analogous way.  Thus considering the 
representation $(\sfrac{1}{2},\sfrac{1}{2})$ of the group $SU(2) \times 
SU(2)$ and denoting $\pm$ the basis vector corresponding to the 
eigenvalues $\pm\sfrac{1}{2}$ of the $J_{3}$ generator in any of the two 
$su(2)$ corresponding algebras, we will assume the following 
``biological'' spin structure:
\begin{eqnarray}
&su(2)_{H}& \nonumber \\
C \equiv (+,+) &\qquad\longleftrightarrow\qquad& U \equiv (-,+) 
\nonumber \\
&& \nonumber \\
su(2)_{V} \updownarrow && \updownarrow su(2)_{V} \\
&& \nonumber \\
G \equiv (+,-) &\qquad\longleftrightarrow\qquad& A \equiv (-,-)
\nonumber \\
&su(2)_{H}& \nonumber
\label{eq1}
\end{eqnarray}
the subscripts $H$ (:= horizontal) and $V$ (:= vertical) being just 
added to specify the group actions.

Now, let us turn our attention towards the representations of ${\cal 
U}_{q \rightarrow 0}({\cal G})$ and more specifically to their crystal 
bases.  In statistical mechanics, the $q \rightarrow 0$ limit of a 
deformed (quantum) algebra can be interpreted as the absolute zero 
temperature in a lattice model.  Introducing in ${\cal U}_{q \rightarrow 
0}({\cal G})$ the operators $\tilde e_{i}$ and $\tilde f_{i}$ 
($i=1,\dots,\mbox{rank }{\cal G}$) after modification of the simple root 
vectors $e_{i}$ and $f_{i}$ of ${\cal U}_{q}({\cal G})$, a particular 
kind of basis in a ${\cal U}_{q}({\cal G})$-module can be defined.  Such 
a basis is called a crystal basis and carries the property to undergo in 
a specially simple way the action of the $\tilde e_{i}$ and $\tilde 
f_{i}$ operators: as an example, for any couple of vectors $u,v$ in the 
crystal basis ${\cal B}$, one gets $u=\tilde e_{i}v$ if and only if 
$v=\tilde f_{i}u$.  One must note that there is no objection to consider 
the four states C, U, G, A defined in (\ref{eq1}) as constituting a 
crystal basis for the $(\sfrac{1}{2},\sfrac{1}{2})$ module of ${\cal 
U}_{q \rightarrow 0}(sl(2) \oplus sl(2))$.  More interesting for our 
purpose is the crystal basis in the tensorial product of two 
representations.  Then the following theorem holds \cite{Kashi}:

\medskip

\noindent 
Let ${\cal B}_{1}$ and ${\cal B}_{2}$ be the crystal bases of the 
$M_{1}$ and $M_{2}$ ${\cal U}_{q \rightarrow 0} ({\cal G})$-modules 
respectively.  Then for $u \in {\cal B}_{1}$ and $v \in {\cal B}_{2}$, 
we have:
\begin{eqnarray}
&& \tilde f_{i}(u \otimes v) = \left\{
\begin{array}{ll}
\tilde f_{i}u \otimes v & \exists \, n \ge 1 \mbox{ such that }
\tilde f_{i}^nu \ne 0 \mbox{ and } \tilde e_{i}v = 0 \\
u \otimes \tilde f_{i}v & \mbox{otherwise} \\
\end{array} \right. \\
&& \tilde e_{i}(u \otimes v) = \left\{
\begin{array}{ll}
u \otimes \tilde e_{i}v & \exists \, n \ge 1 \mbox{ such that }
\tilde e_{i}^nv \ne 0 \mbox{ and } \tilde f_{i}u = 0 \\
\tilde e_{i}u \otimes v & \mbox{otherwise} \\
\end{array} \right.
\end{eqnarray}
To represent a codon, we will have to perform the tensor product of 
three $(\sfrac{1}{2},\sfrac{1}{2})$ representations of ${\cal U}_{q 
\rightarrow 0}(sl(2) \oplus sl(2))$ .  However, it is well-known -- and 
easy to check from Tables \ref{table1}, \ref{table2} -- that in a multiplet 
of codons relative to a specific amino-acid, the two first bases 
constituent of a codon are ``relatively stable'', the degeneracy being 
mainly generated by the third nucleotide.  For that reason, we will 
prefer to examine a codon as a 2+1 state instead of a simple triplet.  
So, let us consider in detail the first tensor product:
\begin{equation}
(\sfrac{1}{2},\sfrac{1}{2}) \, \otimes \, (\sfrac{1}{2},\sfrac{1}{2}) 
= (1,1) \, \oplus \, (1,0) \, \oplus \, (0,1) \, \oplus \, (0,0) 
\label{eq4}
\end{equation}
where inside the parenthesis, $j=0,\sfrac{1}{2},1$ is put in place of 
the $2j+1=1,2,3$ respectively dimensional $SU(2)$ representation.  
We get, using Theorem 1, the following tableau:
\[
\begin{array}{lcccc}
\rightarrow \,\, su(2)_H \qquad \qquad &(0,0) & (CA) &\qquad \qquad 
(1,0) & (\begin{array}{ccc} CG & UG & UA \end{array}) \\
\downarrow \\
su(2)_V &(0,1) & \left( \begin{array}{c} CU \\ GU \\ GA \end{array} 
\right) & \qquad \qquad (1,1) & \left( \begin{array}{ccc}
CC & UC & UU \\ GC & AC & AU \\ GG & AG & AA \\ \end{array}Ê\right)
\end{array}	
\]
{From} Tables \ref{table1} and \ref{table2}, the dinucleotide states 
formed by the first two nucleotides in a codon can be put in 
correspondence with quadruplets, doublets or singlets of codons relative 
to an amino-acid.  Note that the sextets (resp.  triplets) are viewed as 
the sum of a quadruplet and a doublet (resp.  a doublet and a singlet).  
The dinucleotide states associated to the quadruplets (as well as those 
included in the sextets) of codons satisfy:
\begin{equation}
J_{H,3}^d > 0 \quad \mbox{or} \quad J_{H,3}^d = 0 \,, \ \ J_{V,3}^d \ge 0 \,, 
\ \ J_{V}^d \ne 0 \,.
\label{eqA}
\end{equation}
where $J_{H,3}^d$ and $J_{V,3}^d$ are the third components of the spin 
generators of the dinucleotide states.   \\
The dinucleotide states associated to the doublets (as well as those 
included in the triplets) and eventually to the singlets of codons are 
such that:
\begin{equation}
J_{H,3}^d < 0 \quad \mbox{or} \quad J_{H,3}^d = 0 \,, \ \ J_{V,3}^d < 0 
\mbox{ or } J_{V}^d = 0 \,.
\label{eqB}
\end{equation}
On the other hand, if we consider the three-fold tensor product, the 
content into irreducible representations of ${\cal U}_{q \rightarrow 
0}(sl(2) \oplus sl(2))$ is given by:
\begin{equation}
(\sfrac{1}{2},\sfrac{1}{2}) \otimes (\sfrac{1}{2},\sfrac{1}{2}) 
\otimes (\sfrac{1}{2},\sfrac{1}{2}) = (\sfrac{3}{2},\sfrac{3}{2}) 
\oplus 2 \, (\sfrac{3}{2},\sfrac{1}{2}) \oplus 2 \, 
(\sfrac{1}{2},\sfrac{3}{2}) \oplus 4 \, (\sfrac{1}{2},\sfrac{1}{2})
\label{eq5}
\end{equation}
The structure of the irreducible representations of the r.h.s. of Eq.  
(\ref{eq5}) is:
\begin{eqnarray*}
&(\sfrac{3}{2},\sfrac{3}{2})  \equiv \left( \begin{array}{cccc} 
CCC & UCC & UUC & UUU \\ GCC & ACC & AUC & AUU \\
GGC & AGC & AAC & AAU \\ GGG & AGG & AAG & AAA \\
\end{array} \right)& \\
&& \\
&(\sfrac{3}{2},\sfrac{1}{2}) \equiv \left( \begin{array}{cccc} 
CCG & UCG & UUG & UUA \\ GCG & ACG & AUG & AUA \\
\end{array} \right)& \\ 
&& \\
&(\sfrac{3}{2},\sfrac{1}{2})' \equiv \left( \begin{array}{cccc} 
CGC & UGC & UAC & UAU \\ CGG & UGG & UAG & UAA \\ 
\end{array} \right)& \\
&& \\
&(\sfrac{1}{2},\sfrac{3}{2}) \equiv \left( \begin{array}{cc}
CCU & UCU \\ GCU & ACU \\ GGU & AGU \\ GGA & AGA \\ 
\end{array} \right) \qquad
(\sfrac{1}{2},\sfrac{3}{2})' \equiv \left( \begin{array}{cc}
CUC & CUU \\ GUC & GUU \\ GAC & GAU \\ GAG & GAA \\ 
\end{array} \right)& \\
&& \\
&(\sfrac{1}{2},\sfrac{1}{2}) \equiv \left( \begin{array}{cc}
CCA & UCA \\ GCA & ACA \\ \end{array} \right) \qquad
(\sfrac{1}{2},\sfrac{1}{2})' \equiv \left( \begin{array}{cc}
CGU & UGU \\ CGA &UGA \\ \end{array} \right)& \\
&& \\
&(\sfrac{1}{2},\sfrac{1}{2})'' \equiv \left( \begin{array}{cc}
CUG & CUA \\ GUG & GUA \\ \end{array} \right) \qquad
(\sfrac{1}{2},\sfrac{1}{2})''' \equiv \left( \begin{array}{cc}
CAC & CAU \\ CAG & CAA \end{array} \right)&
\end{eqnarray*}

\medskip

As expected from formulae (\ref{eqA}) and (\ref{eqB}), our model cannot 
gather codons associated to one particular amino-acid in the same 
irreducible multiplet. However, it is possible to construct an 
operator ${\cal R}$ out of the algebra ${\cal U}_{q \rightarrow 0}(sl(2) 
\oplus sl(2))$, acting on the codons, that will describe the genetic 
code in the following way:

{\em Two codons have the same eigenvalue under  ${\cal R}$ if and only if 
they are associated to the same amino-acid.}

This operator will be called the {\em reading operator}.  It has the 
following form:
\begin{eqnarray}
{\cal R} &=& \sfrac{4}{3} c_{1} \, C_{H} + \sfrac{4}{3} c_{2} \, C_{V} 
- 4c_{1} \, {\cal P}_{1} \, J_{H,3} - 4c_{2} \, {\cal P}_{2} \, J_{V,3} 
+ ({\cal P}_{3} \, c_{3} + {\cal P}_{4} \, c_{4}) \, J_{V,3} \nonumber \\
&& + {\cal P}_{5} \, c_{5} \, (\sfrac{1}{2}-J_{V,3}^{(3)}) 
+ ({\cal P}_{6} \, q + {\cal P}'_{6} \, q') \, 
(\sfrac{1}{2}-J_{V,3}^{(3)}) \, J_{H,3}^{(3)} \,.
\label{eq6}
\end{eqnarray}
In Eq.  (\ref{eq6}), the operators $J_{H,3}$ and $J_{V,3}$ are the third 
components of the total spin generators of the algebra ${\cal U}_{q 
\rightarrow 0}(sl(2) \oplus sl(2))$, $J_{H,3}^{(3)}$, $J_{V,3}^{(3)}$ 
are the third components corresponding to the third nucleotide of a 
codon.  Of course, these last two operators can be replaced by 
$J_{\alpha,3}^{(3)} = J_{\alpha,3} - J_{\alpha,3}^d$ ($\alpha = H,V$).  \\
The operator $C_{\alpha}$ ($\alpha = H,V$) is a ``Casimir'' operator of 
${\cal U}_{q \rightarrow 0}(sl(2))$ in the crystal basis.  It is 
characterized by the property that it commutes with $J_{\pm,H}, 
J_{\pm,V}$ and $J_{H,3}, J_{V,3}$ (where $J_{\pm,H}, J_{\pm,V}$ are the 
generators with a well-defined behaviour for $q \rightarrow 0$) and its 
eigenvalues on any vector basis of an irreducible representation of 
highest weight $J$ is $J(J+1)$, i.e.  the same as the undeformed 
standard second degree Casimir operator of $sl(2)$.  Its explicit 
expression is
\begin{equation}
C = (J_3)^{2} + \frac12 \sum_{n \in \ZZ_+}
\sum_{k=0}^n (J_-)^{n-k} (J_+)^n (J_-)^k \,.
\end{equation}
Note that for $sl(2)_{q \rightarrow 0}$ the ``Casimir'' operator is an 
infinite series of powers of $J_-$ and $J_+$.  However in any finite 
irreducible representation only a finite number of terms gives a 
non-vanishing contribution.  \\
${\cal P}_{i}$ ($i=1,\dots,5$) are projectors given by the following 
expressions:
\begin{eqnarray}
&&  {\cal P}_{1} = J_{H+}^d \, J_{H-}^d \,, \nonumber \\
&&  {\cal P}_{2} = J_{V+}^d \, J_{V-}^d \,, \nonumber \\
&&  {\cal P}_{3} = J_{H-}^d \, J_{H+}^d 
(2-J_{H+}^d \, J_{H-}^d-J_{V+}^d \, J_{V-}^d) + 
(1-J_{H-}^d \, J_{H+}^d)(1-J_{H+}^d \, J_{H-}^d)
(1-J_{V+}^d \, J_{V-}^d) \,, \nonumber \\
&& {\cal P}_{4} = (J_{H-}^d \, J_{H+}^d) \,\, [(J_{H+}^d \, J_{H-}^d)
(1-J_{V+}^d \, J_{V-}^d) + (J_{V+}^d \, J_{V-}^d)(J_{V-}^d \, J_{V+}^d)
(1-J_{H+}^d \, J_{H-}^d )] \,, \nonumber \\
&&  {\cal P}_{5} =  (J_{H-}^d \, J_{H+}^d)(J_{V-}^d \, J_{V+}^d)
(J_{H+}^d \, J_{H-}^d)(1-J_{V+}^d \, J_{V-}^d) \,.
\label{eq7}
\end{eqnarray}
The projectors ${\cal P}_6,{\cal P}'_6$ appear only for the eukariotic 
code. Their expressions are given by:
\begin{eqnarray}
&&  {\cal P}_{6} = (J_{H-}^d \, J_{H+}^d)(J_{V-}^d \, J_{V+}^d)
(1-J_{H+}^d \, J_{H-}^d)(J_{V+}^d \, J_{V-}^d) \,, \nonumber \\
&&  {\cal P}'_{6} = (J_{H-}^d \, J_{H+}^d)(1-J_{V-}^d \, J_{V+}^d)
(J_{H+}^d \, J_{H-}^d)(1-J_{V+}^d \, J_{V-}^d) \,.
\label{eq7bis}
\end{eqnarray}
The terms in $c_{1}$ and $c_{2}$ are responsible for the structure in 
quadruplets (given essentially by the dinucleotide content).  The terms 
in $c_{3}$ give rise to the splitting of the quadruplets into doublets.  
The terms in $c_{4}$ and $c_{5}$ lead to the sextets.  Finally, the 
terms in $q$ and $q'$, that appear only in the eukariotic code, are 
responsible for the singlet and triplet structure.

Now, using the values of the quantum numbers $J_{H}$, $J_{V}$, 
$J_{H,3}$, $J_{V,3}$, $J_{\alpha\pm}^d \, J_{\alpha\pm}^d$ 
($\alpha=H,V$) of the codons given in Tables 3 and 4, one can compute 
the action of the reading operator ${\cal R}$ on each of the 64 codons.

Although the eukariotic code (EC) seems to be a universal genetic code, 
it appears in some way as an advanced form of the vertebral 
mitochondrial code (VMC).  Indeed there is very few difference between 
the two codes.  The codons in the VMC are organized into 2 sextets, 6 
quadruplets and 14 doublets.  When evolving from the VMC to the EC, one 
doublet and one quartet merge together to form a sextet while two other 
doublets split into four singlets, two of them gluing with existing 
doublets to form two triplets.  The final result for the EC is 3 
sextets, 5 quadruplets, 10 doublets, 2 triplets and 2 singlets.  Hence, 
it appears natural to start to calculate ${\cal R}$ for the vertebral 
mitochondrial code.

\medskip

\noindent
\underline{{\bf a) Vertebral Mitochondrial Code:}}  \\
One finds the following eigenvalues of the reading operator 
${\cal R}$ in the case of the vertebral mitochondrial code, identifying 
the amino-acids with its corresponding codons
(Ser corresponds to the codons UCX (X=C,U,G,A) while Ser$'$ corresponds to 
the codons AGC/AGU; similarly Leu is related to the quartet CUX and 
Leu$'$ to the doublet UUG/UUA; finally, Arg is given by the quartet CGX 
and Ter$'$ to the doublet AGG/AGA):  \\
\begin{equation}
\begin{array}{ll}
\bigg. \mbox{Pro} = - c_{1} - c_{2} & \qquad
\bigg. \mbox{Thr} = 3 c_{1} + 3 c_{2} \\
\bigg. \mbox{Ala} = - c_{1} + 3 c_{2} & \qquad
\bigg. \mbox{Ser} = 3 c_{1} - c_{2} \\
\bigg. \mbox{Asp} = c_{1} + 5 c_{2} - \sfrac{1}{2} c_{3} & \qquad
\bigg. \mbox{Glu} = c_{1} + 5 c_{2} - \sfrac{3}{2} 3c_{3} \\
\bigg. \mbox{Tyr} = 5 c_{1} + c_{2} + \sfrac{1}{2} c_{3} & \qquad
\bigg. \mbox{Ter} = 5 c_{1} + c_{2} - \sfrac{1}{2} c_{3} \\
\bigg. \mbox{Asn} = 5 c_{1} + 5 c_{2} - \sfrac{1}{2} c_{3} & \qquad
\bigg. \mbox{Lys} = 5 c_{1} + 5 c_{2} - \sfrac{3}{2} c_{3} \\
\bigg. \mbox{His} = c_{1} + c_{2} + \sfrac{1}{2} c_{3} & \qquad
\bigg. \mbox{Gln} = c_{1} + c_{2} - \sfrac{1}{2} c_{3} \\
\bigg. \mbox{Arg} = - c_{1} + c_{2} & \qquad
\bigg. \mbox{Gly} = - c_{1} + 5 c_{2} \\
\bigg. \mbox{Cys} = 3 c_{1} + c_{2} + \sfrac{1}{2} c_{3} + 
\sfrac{1}{2} c_{4} & \qquad
\bigg. \mbox{Trp} = 3 c_{1} + c_{2} - \sfrac{1}{2} c_{3} - 
\sfrac{1}{2} c_{4} \\
\bigg. \mbox{Ser$'$} = 3 c_{1} + 5 c_{2} - \sfrac{1}{2} c_{3} - 
\sfrac{1}{2} c_{4} & \qquad
\bigg. \mbox{Ter$'$} = 3 c_{1} + 5 c_{2} - \sfrac{3}{2} c_{3} - 
\sfrac{3}{2} c_{4} + c_{5} \\
\bigg. \mbox{Val} = c_{1} + 3 c_{2} & \qquad
\bigg. \mbox{Leu} = c_{1} - c_{2} \\
\bigg. \mbox{Phe} = 5 c_{1} - c_{2} + \sfrac{3}{2} c_{3} & \qquad
\bigg. \mbox{Leu$'$} = 5 c_{1} - c_{2} + \sfrac{1}{2} c_{3} \\
\bigg. \mbox{Ile} = 5 c_{1} + 3 c_{2} + \sfrac{1}{2} c_{3} + 
\sfrac{1}{2} c_{4} & \qquad
\bigg. \mbox{Met} = 5 c_{1} + 3 c_{2} - \sfrac{1}{2} c_{3} - 
\sfrac{1}{2} c_{4} \\
\end{array}
\label{eqvalvmc}
\end{equation}
The parameters $c_{3}$, $c_{4}$ are fixed by the following requirements.  
The condition Leu = Leu$'$ leads to the expression of the coefficient 
$c_{3}$ in function of $c_{1}$ and $c_{2}$, and one obtains $c_{3} = 
-8c_{1}$.  At this point, one is led to add a correcting term in ${\cal 
R}$ since the symmetry of the genetic code implies Ile = Val and Cys = 
Arg as soon as Leu = Leu$'$ while Ser$'$ is not equal to Ser.  Hence the 
projector ${\cal P}_{4}$ has a non-vanishing value on the $AG$, $UG$ and 
$AU$ dinucleotides.  The condition Ser$'$ = Ser then implies $c_{4} = 8 
c_{1} + 12 c_{2}$.  At this point, Ile and Val on the one hand, and Cys 
and Arg on the other hand become different as required.  Finally, the 
parameter $c_{5}$ is fixed for the VMC by requiring that Ter$'$ = Ter.  
One finds $c_{5} = 6c_{1}+14c_{2}$.  The demand to be satisfied by 
${\cal R}$ in order to provide different eigenvalues to codons 
associated to different amino-acids implies the non-vanishing of $c_{1}$ 
and $c_{2}$.  This leads, after a rescaling, to express the reading 
operator for the vertebral mitochondrial code as (where $c \equiv 
c_{1}/c_{2}$):
\begin{eqnarray}
{\cal R}_{VMC}(c) &=& \sfrac{4}{3} c \, C_{H} + \sfrac{4}{3} \, C_{V} 
- 4c \, {\cal P}_{1} \, J_{H,3} - 4 \, {\cal P}_{2} \, J_{V,3}  
+ (-8c \, {\cal P}_{3} + (8c+12) \, {\cal P}_{4}) \, J_{V,3}
\nonumber \\
&& + (6c+14) \, {\cal P}_{5} \, (\sfrac{1}{2}-J_{V,3}^{(3)}) \,.
\label{eq8}
\end{eqnarray}
and therefore to the following values for the amino-acids:
\begin{equation}
\begin{array}{c|c||c|c||c|c}
\mbox{a.a.} & \mbox{value of the codon} &
\mbox{a.a.} & \mbox{value of the codon} &
\mbox{a.a.} & \mbox{value of the codon} \\
\hline
\bigg. \mbox{Ala} & - c + 3 &
\bigg. \mbox{Gly} & - c + 5 &
\bigg. \mbox{Pro} & - c - 1 \\
\bigg. \mbox{Arg} & - c + 1 & 
\bigg. \mbox{His} & -3 c + 1 &
\bigg. \mbox{Ser} & 3 c - 1 \\
\bigg. \mbox{Asn} & 9 c + 5 &
\bigg. \mbox{Ile} & 5 c + 9 &
\bigg. \mbox{Thr} & 3 c + 3 \\
\bigg. \mbox{Asp} & 5 c + 5 &
\bigg. \mbox{Leu} & c - 1 & 
\bigg. \mbox{Trp} & 3 c - 5 \\
\bigg. \mbox{Cys} & 3 c + 7 &
\bigg. \mbox{Lys} & 17 c + 5 & 
\bigg. \mbox{Tyr} & c + 1 \\
\bigg. \mbox{Gln} & 5 c + 1 &
\bigg. \mbox{Met} & 5 c - 3 & 
\bigg. \mbox{Val} & c + 3 \\
\bigg. \mbox{Glu} & 13 c + 5 &
\bigg. \mbox{Phe} & -7 c - 1 & 
\bigg. \mbox{Ter} & 9 c + 1 \\
\end{array}
\end{equation}
\centerline{The vertebral mitochondrial code}

We remark that the reading operator ${\cal R}_{VMC}(c)$ can be used for 
any real value of $c$, except those confering the same eigenvalue to 
codons relative to two different amino-acids.  These forbidden values 
are the following: $-7$, $-5$, $-4$, $-3$, $-\sfrac{5}{2}$, 
$-\sfrac{7}{3}$, $-2$, $-\sfrac{5}{3}$, $-\sfrac{3}{2}$, 
$-\sfrac{4}{3}$, $-1$, $-\sfrac{5}{6}$, $-\sfrac{4}{5}$, 
$-\sfrac{3}{4}$, $-\sfrac{5}{7}$, $-\sfrac{2}{3}$, $-\sfrac{3}{5}$, 
$-\sfrac{1}{2}$, $-\sfrac{3}{7}$, $-\sfrac{2}{5}$, $-\sfrac{3}{8}$, 
$-\sfrac{1}{3}$, $-\sfrac{3}{10}$, $-\sfrac{2}{7}$, $-\sfrac{1}{4}$, 
$-\sfrac{2}{9}$, $-\sfrac{1}{5}$, $-\sfrac{1}{6}$, $-\sfrac{1}{7}$, 
$-\sfrac{1}{8}$, $-\sfrac{1}{9}$, $0$, $\sfrac{1}{7}$, $\sfrac{1}{6}$, 
$\sfrac{1}{5}$, $\sfrac{1}{4}$, $\sfrac{1}{3}$, $\sfrac{2}{5}$, 
$\sfrac{1}{2}$, $\sfrac{2}{3}$, $1$, $\sfrac{4}{3}$, $\sfrac{3}{2}$, 
$2$, $\sfrac{5}{2}$, $3$, $4$, $5$.

\bigskip

\noindent
\underline{{\bf b) The Eukariotic Code:}}  \\
In the case of the eukariotic code, most of the eigenvalues of the reading 
operator are the same. The difference between VMC and EC comes i) from the 
doublets Met and Trp that split into singlets Met (AUG) + Ile$''$ (AUA) 
and Trp (UGG) + Ter$''$ (UGA), and ii) from the doublet Ter$'$ that 
merge with the quartet Arg to form a sextet.
The eigenvalues for the new structures are the following:
\begin{equation}
\begin{array}{l}
\bigg. \mbox{Ter}'' = 3 c_{1} + c_{2} - \sfrac{1}{2} c_{3} - 
\sfrac{1}{2} c_{4} - q' \\
\bigg. \mbox{Trp} = 3 c_{1} + c_{2} - \sfrac{1}{2} c_{3} - 
\sfrac{1}{2} c_{4} + q' \\
\bigg. \mbox{Ile}'' = 5 c_{1} + 3 c_{2} - \sfrac{1}{2} c_{3} - 
\sfrac{1}{2} c_{4} - q \\
\bigg. \mbox{Met} = 5 c_{1} + 3 c_{2} - \sfrac{1}{2} c_{3} - 
\sfrac{1}{2} c_{4} + q \\
\end{array}
\end{equation}
The parameters $c_{3}$, $c_{4}$ are given as in the VMC. The parameter 
$c_{5}$ is now fixed by the condition Ter$'$ = Arg.  One obtains $c_{5} 
= -4c_{1}+14c_{2}$.  The parameters $q$ and $q'$ describe the splitting 
of the doublets Met and Trp into the singlets: they are determined by 
requiring Ile$''$ = Ile and Ter$''$ = Ter.  Il follows that $q = 
-12c_{2}$ and $q' = -6c_{1}-6c_{2}$.  Hence the reading operator for the 
eukariotic code reads as:
\begin{eqnarray}
{\cal R}_{EC}(c) &=& \sfrac{4}{3} c \, C_{H} + \sfrac{4}{3} \, C_{V} 
- 4c \, {\cal P}_{1} \, J_{H,3} - 4 \, {\cal P}_{2} \, J_{V,3} 
+ (- 8c \, {\cal P}_{3} + (8c+12) \, {\cal P}_{4}) \, J_{V,3} \nonumber \\
&&+ (-4c+14) \, {\cal P}_{5} \, (\sfrac{1}{2}-J_{V,3}^{(3)})
- 6 \, (2 \, {\cal P}_{6} +  (c + 1) \, {\cal P}'_{6}) \, 
(\sfrac{1}{2}-J_{V,3}^{(3)}) \, J_{H,3}^{(3)} \,.
\label{eq9}
\end{eqnarray}
where as in case (a) we have achieved a rescaling and $c \equiv 
c_{1}/c_{2}$. This leads to the following values for the amino-acids:
\begin{equation}
\begin{array}{c|c||c|c||c|c}
\mbox{a.a.} & \mbox{value of the codon} &
\mbox{a.a.} & \mbox{value of the codon} &
\mbox{a.a.} & \mbox{value of the codon} \\
\hline
\bigg. \mbox{Ala} & - c + 3 &
\bigg. \mbox{Gly} & - c + 5 &
\bigg. \mbox{Pro} & - c - 1 \\
\bigg. \mbox{Arg} & - c + 1 &
\bigg. \mbox{His} & -3 c + 1 &
\bigg. \mbox{Ser} & 3 c - 1 \\
\bigg. \mbox{Asn} & 9 c + 5 &
\bigg. \mbox{Ile} & 5 c + 9 &
\bigg. \mbox{Thr} & 3 c + 3 \\
\bigg. \mbox{Asp} & 5 c + 5 &
\bigg. \mbox{Leu} & c - 1 &
\bigg. \mbox{Trp} & -3 c - 11 \\
\bigg. \mbox{Cys} & 3 c + 7 &
\bigg. \mbox{Lys} & 17 c + 5 &
\bigg. \mbox{Tyr} & c + 1 \\
\bigg. \mbox{Gln} & 5 c + 1 &
\bigg. \mbox{Met} & 5 c - 15 &
\bigg. \mbox{Val} & c + 3 \\
\bigg. \mbox{Glu} & 13 c + 5 &
\bigg. \mbox{Phe} & -7 c - 1 &
\bigg. \mbox{Ter} & 9 c + 1 \\
\end{array}
\end{equation}
\centerline{The eukariotic code}

As in the case (a), we have to rule out the values of the parameter $c$ 
such that different amino-acids get the same eigenvalue under ${\cal 
R}_{EC}(c)$.  The forbidden values now are the following: $-8$, $-7$, 
$-6$, $-5$, $-4$, $-\sfrac{7}{2}$, $-3$, $-\sfrac{5}{2}$, 
$-\sfrac{7}{3}$, $-2$, $-\sfrac{5}{3}$, $-\sfrac{3}{2}$, 
$-\sfrac{4}{3}$, $-1$, $-\sfrac{5}{6}$, $-\sfrac{4}{5}$, 
$-\sfrac{3}{4}$, $-\sfrac{2}{3}$, $-\sfrac{3}{5}$, $-\sfrac{1}{2}$, 
$-\sfrac{3}{7}$, $-\sfrac{2}{5}$, $-\sfrac{3}{8}$, $-\sfrac{1}{3}$, 
$-\sfrac{3}{10}$, $-\sfrac{2}{7}$, $-\sfrac{1}{4}$, $-\sfrac{2}{9}$, 
$-\sfrac{1}{5}$, $-\sfrac{1}{6}$, $-\sfrac{1}{7}$, $-\sfrac{1}{8}$, 
$-\sfrac{1}{9}$, $0$, $\sfrac{1}{7}$, $\sfrac{1}{5}$, $\sfrac{1}{4}$, 
$\sfrac{1}{3}$, $\sfrac{2}{5}$, $\sfrac{1}{2}$, $\sfrac{2}{3}$, $1$, 
$\sfrac{7}{6}$, $\sfrac{3}{2}$, $2$, $\sfrac{7}{3}$, $\sfrac{5}{2}$, 
$\sfrac{8}{3}$, $3$, $\sfrac{10}{3}$, $\sfrac{7}{2}$, $4$, 
$\sfrac{9}{2}$, $7$, $9$, $11$.

\bigskip

The simple model that we propose needs obviously to be developed. First 
on the symmetry point of view, it would be nice to understand or at 
least to include naturally in our approach the existence of sextets. 
Concerning the group structure, we have chosen ${\cal U}_{q \rightarrow 
0}({\cal G})$ with a minimal group $G = SU(2) \times SU(2)$, keeping in 
mind a physical interpretation. Of course, a larger symmetry might be 
of some help. As a second step, it will be reasonable to consider a more 
realistic model including interactions among bases. We wish to be soon 
able to apply our approach on the one hand for mutations and on the 
other hand in the fundamental problem of genome sequence.

Let us end this note by the following general remark. There are intense 
efforts these days to develop an interface between physics and biology. 
Different approaches are considered, among them the study of the DNA as 
an ideal polymer in the framework of statistical physics. But no direct 
connection between biology and elementary particle physics already 
showed up, in our knowledge. We hope that our proposal will raise up the 
interest of elementary particle physics in biology.

\bigskip

{\bf Acknowledgements}

One of us (P.S.) is indebted to G. Brun and J.L. Darlix, biologists in 
ENS Lyon, for providing with informations and for encouragements, to A. 
Figureau for useful discussions.  The authors wish also to thank D. 
Arnaudon for mentioning to us ref.  \cite{Kashi}.

\newpage

\begin{table}[t]
\centering
\begin{tabular}{|cc||cc||cc||cc|}
\hline
CCC & Pro & UCC & Ser & GCC & Ala & ACC & Thr \\
CCU & Pro & UCU & Ser & GCU & Ala & ACU & Thr \\
CCG & Pro & UCG & Ser & GCG & Ala & ACG & Thr \\
CCA & Pro & UCA & Ser & GCA & Ala & ACA & Thr \\
\hline
CUC & Leu & UUC & Phe & GUC & Val & AUC & Ile \\
CUU & Leu & UUU & Phe & GUU & Val & AUU & Ile \\
CUG & Leu & UUG & Leu & GUG & Val & AUG & Met \\
CUA & Leu & UUA & Leu & GUA & Val & AUA & Met \\
\hline
CGC & Arg & UGC & Cys & GGC & Gly & AGC & Ser \\
CGU & Arg & UGU & Cys & GGU & Gly & AGU & Ser \\
CGG & Arg & UGG & Trp & GGG & Gly & AGG & Ter \\
CGA & Arg & UGA & Trp & GGA & Gly & AGA & Ter \\
\hline
CAC & His & UAC & Tyr & GAC & Asp & AAC & Asn \\
CAU & His & UAU & Tyr & GAU & Asp & AAU & Asn \\
CAG & Gln & UAG & Ter & GAG & Glu & AAG & Lys \\
CAA & Gln & UAA & Ter & GAA & Glu & AAA & Lys \\
\hline
\end{tabular}
\caption{The vertebral mitochondrial code.\label{table1}} 
\end{table}

\begin{table}[b]
\centering
\begin{tabular}{|cc||cc||cc||cc|}
\hline
CCC & Pro & UCC & Ser & GCC & Ala & ACC & Thr \\
CCU & Pro & UCU & Ser & GCU & Ala & ACU & Thr \\
CCG & Pro & UCG & Ser & GCG & Ala & ACG & Thr \\
CCA & Pro & UCA & Ser & GCA & Ala & ACA & Thr \\
\hline
CUC & Leu & UUC & Phe & GUC & Val & AUC & Ile \\
CUU & Leu & UUU & Phe & GUU & Val & AUU & Ile \\
CUG & Leu & UUG & Leu & GUG & Val & AUG & Met \\
CUA & Leu & UUA & Leu & GUA & Val & AUA & Ile \\
\hline
CGC & Arg & UGC & Cys & GGC & Gly & AGC & Ser \\
CGU & Arg & UGU & Cys & GGU & Gly & AGU & Ser \\
CGG & Arg & UGG & Trp & GGG & Gly & AGG & Arg \\
CGA & Arg & UGA & Ter & GGA & Gly & AGA & Arg \\
\hline
CAC & His & UAC & Tyr & GAC & Asp & AAC & Asn \\
CAU & His & UAU & Tyr & GAU & Asp & AAU & Asn \\
CAG & Gln & UAG & Ter & GAG & Glu & AAG & Lys \\
CAA & Gln & UAA & Ter & GAA & Glu & AAA & Lys \\
\hline
\end{tabular}
\caption{The eukariotic code.\label{table2}} 
\end{table}

\clearpage
\begin{table}[htbp]
\centering
\begin{tabular}{|cc||cc|cc||cc||cc|cc|}
\hline
codon & a.a. & $J_H$ & $J_V$ & $J_{H,3}$ & $J_{V,3}$ &
codon & a.a. & $J_H$ & $J_V$ & $J_{H,3}$ & $J_{V,3}$  \\
\hline
CCC & Pro & 3/2 & 3/2 & 3/2 & 3/2 & GCC & Ala & 3/2 & 3/2 & 3/2 & 1/2  \\ 
CCU & Pro & 1/2 & 3/2 & 1/2 & 3/2 & GCU & Ala &  1/2 & 3/2 & 1/2 & 1/2  \\
CCG & Pro & 3/2 & 1/2 & 3/2 & 1/2 & GCG & Ala & 3/2 & 1/2 & 3/2 & -1/2  \\
CCA & Pro & 1/2 & 1/2 & 1/2 & 1/2 & GCA & Ala & 1/2 & 1/2 & 1/2 & -1/2  \\
\hline
CUC & Leu & 1/2 & 3/2 & 1/2 & 3/2 & GUC & Val & 1/2 & 3/2 & 1/2 & 1/2  \\
CUU & Leu & 1/2 & 3/2 & -1/2 & 3/2 & GUU & Val & 1/2 & 3/2 & -1/2 & 1/2  \\
CUG & Leu & 1/2 & 1/2 & 1/2 & 1/2 & GUG & Val & 1/2 & 1/2 & 1/2 & -1/2  \\
CUA & Leu & 1/2 & 1/2 & -1/2 & 1/2 & GUA & Val & 1/2 & 1/2 & -1/2 & -1/2  \\
\hline
CGC & Arg &  3/2 & 1/2 & 3/2 & 1/2 & GGC & Gly & 3/2 & 3/2 & 3/2 & -1/2  \\
CGU & Arg & 1/2 & 1/2 & 1/2 & 1/2 & GGU & Gly & 1/2 & 3/2 & 1/2 & -1/2  \\
CGG & Arg &  3/2 & 1/2 & 3/2 & -1/2 & GGG & Gly & 3/2 & 3/2 & 3/2 & -3/2  \\
CGA & Arg &  1/2 & 1/2 & 1/2 & -1/2 & GGA & Gly & 1/2 & 3/2 & 1/2 & -3/2  \\
\hline
CAC & His & 1/2 & 1/2 & 1/2 & 1/2 & GAC & Asp & 1/2 & 3/2 & 1/2 & -1/2  \\
CAU & His & 1/2 & 1/2 & -1/2 & 1/2 & GAU & Asp & 1/2 & 3/2 & -1/2 & -1/2  \\
CAG & Gln & 1/2 & 1/2 & 1/2 & -1/2 & GAG & Glu & 1/2 & 3/2 & 1/2 & -3/2  \\
CAA & Gln & 1/2 & 1/2 & -1/2 & -1/2 & GAA & Glu & 1/2 & 3/2 & -1/2 & -3/2  \\
\hline
UCC & Ser & 3/2 & 3/2 & 1/2 & 3/2 & ACC & Thr &  3/2 & 3/2 & 1/2 & 1/2  \\
UCU & Ser & 1/2 & 3/2 & -1/2 & 3/2 & ACU & Thr & 1/2 & 3/2 & -1/2 & 1/2  \\
UCG & Ser & 3/2 & 1/2 & 1/2 & 1/2 & ACG & Thr &  3/2 & 1/2 & 1/2 & -1/2  \\
UCA & Ser & 1/2 & 1/2 & -1/2 & 1/2 & ACA & Thr &  1/2 & 1/2 & -1/2 & -1/2  \\
\hline
UUC & Phe & 3/2 & 3/2 & -1/2 & 3/2 & AUC & Ile & 3/2 & 3/2 & -1/2 & 1/2  \\
UUU & Phe & 3/2 & 3/2 & -3/2 & 3/2 & AUU & Ile & 3/2 & 3/2 & -3/2 & 1/2  \\
UUG & Leu & 3/2 & 1/2 & -1/2 & 1/2 & AUG & Met & 3/2 & 1/2 & -1/2 & -1/2  \\
UUA & Leu & 3/2 & 1/2 & -3/2 & 1/2 & AUA & Met/Ile & 3/2 & 1/2 & -3/2 & -1/2  \\
\hline
UGC & Cys & 3/2 & 1/2 & 1/2 & 1/2 & AGC & Ser & 3/2 & 3/2 & 1/2 & -1/2  \\
UGU & Cys & 1/2 & 1/2 & -1/2 & 1/2 & AGU & Ser & 1/2 & 3/2 & -1/2 & -1/2  \\
UGG & Ter/Trp & 3/2 & 1/2 & 1/2 & -1/2 & AGG & Ter/Arg & 3/2 & 3/2 & 1/2 & -3/2  \\
UGA & Ter & 1/2 & 1/2 & -1/2 & -1/2 & AGA & Ter/Arg &  1/2 & 3/2 & -1/2 & -3/2  \\
\hline
UAC & Tyr & 3/2 & 1/2 & -1/2 & 1/2 & AAC & Asn & 3/2 & 3/2 & -1/2 & -1/2  \\
UAU & Tyr & 3/2 & 1/2 & -3/2 & 1/2 & AAU & Asn & 3/2 & 3/2 & -3/2 & -1/2  \\
UAG & Ter & 3/2 & 1/2 & -1/2 & -1/2 & AAG & Lys & 3/2 & 3/2 & -1/2 & -3/2  \\
UAA & Ter & 3/2 & 1/2 & -3/2 & -1/2 & AAA & Lys & 3/2 & 3/2 & -3/2 & -3/2  \\
\hline
\end{tabular}
\caption{$J_\alpha$, ($J_{\alpha,3}$): values of total/third component 
spin of $su(2)_{\alpha}$, ($\alpha = H,V$).\label{table3}} 
\end{table}
\begin{center}
(in the columns ``amino-acids'' (a.a.), left is for VMC and right for EC)
\end{center}

\clearpage

\begin{table}[t]
\centering
\begin{tabular}{|c||c|c|c|c|}
\hline
dinucl. & $J_{H+}^d \, J_{H-}^d$ & $J_{H-}^d \, J_{H+}^d$ & 
$J_{V+}^d \, J_{V-}^d$ & $J_{V-}^d \, J_{V+}^d$ \\
\hline
CC & 1 & 0 & 1 & 0 \\
CU & 0 & 0 & 1 & 0 \\
CG & 1 & 0 & 0 & 0 \\
CA & 0 & 0 & 0 & 0 \\
\hline
UC & 1 & 1 & 1 & 0 \\
UU & 0 & 1 & 1 & 0 \\
UG & 1 & 1 & 0 & 0 \\
UA & 0 & 1 & 0 & 0 \\
\hline
GC & 1 & 0 & 1 & 1 \\
GU & 0 & 0 & 1 & 1 \\
GG & 1 & 0 & 0 & 1 \\
GA & 0 & 0 & 0 & 1 \\
\hline
AC & 1 & 1 & 1 & 1 \\
AU & 0 & 1 & 1 & 1 \\
AG & 1 & 1 & 0 & 1 \\
AA & 0 & 1 & 0 & 1 \\
\hline
\end{tabular}
\caption{Values of $J_{\alpha,\pm}^d \, J_{\alpha,\pm}^d$ ($\alpha = 
H,V$) for the dinucleotides formed by the first two 
nucleotides.\label{table4}}
\end{table}

\end{document}